\documentclass[twoside,twocolumn,9pt]{article}
\usepackage{extsizes}
\usepackage[super,sort&compress,comma]{natbib} 
\usepackage[version=3]{mhchem}
\usepackage[left=1.5cm, right=1.5cm, top=1.785cm, bottom=2.0cm]{geometry}
\usepackage{balance}
\usepackage{mathptmx}
\usepackage{sectsty}
\usepackage{graphicx} 
\usepackage{lastpage}
\usepackage[format=plain,justification=justified,singlelinecheck=false,font={stretch=1.125,small,sf},labelfont=bf,labelsep=space]{caption}
\usepackage{float}
\usepackage{fancyhdr}
\usepackage{fnpos}
\usepackage[english]{babel}
\addto{\captionsenglish}{%
  
}
\usepackage{array}
\usepackage{droidsans}
\usepackage{charter}
\usepackage[T1]{fontenc}
\usepackage[usenames,dvipsnames]{xcolor}
\usepackage{setspace}
\usepackage[compact]{titlesec}

\definecolor{cream}{RGB}{222,217,201}

\begin{document}

\pagestyle{fancy}
\thispagestyle{plain}
\fancypagestyle{plain}{
\renewcommand{\headrulewidth}{0pt}
}

\makeFNbottom
\makeatletter
\renewcommand\LARGE{\@setfontsize\LARGE{15pt}{17}}
\renewcommand\Large{\@setfontsize\Large{12pt}{14}}
\renewcommand\large{\@setfontsize\large{10pt}{12}}
\renewcommand\footnotesize{\@setfontsize\footnotesize{7pt}{10}}
\makeatother

\renewcommand{\thefootnote}{\fnsymbol{footnote}}
\renewcommand\footnoterule{\vspace*{1pt}%
\color{cream}\hrule width 3.5in height 0.4pt \color{black}\vspace*{5pt}} 
\setcounter{secnumdepth}{5}

\makeatletter 
\renewcommand\@biblabel[1]{#1}            
\renewcommand\@makefntext[1]%
{\noindent\makebox[0pt][r]{\@thefnmark\,}#1}
\makeatother 
\renewcommand{\figurename}{\small{Fig.}~}
\sectionfont{\sffamily\Large}
\subsectionfont{\normalsize}
\subsubsectionfont{\bf}
\setstretch{1.125} 
\setlength{\skip\footins}{0.8cm}
\setlength{\footnotesep}{0.25cm}
\setlength{\jot}{10pt}
\titlespacing*{\section}{0pt}{4pt}{4pt}
\titlespacing*{\subsection}{0pt}{15pt}{1pt}

\fancyfoot{}
\fancyfoot[LO,RE]{\vspace{-7.1pt}}
\fancyfoot[CO]{\vspace{-7.1pt}\hspace{13.2cm}}
\fancyfoot[CE]{\vspace{-7.2pt}\hspace{-14.2cm}}
\fancyfoot[RO]{\footnotesize{\sffamily{ \textbar \hspace{2pt}\thepage}}}
\fancyfoot[LE]{\footnotesize{\sffamily{\thepage~\textbar\hspace{3.45cm} }}}
\fancyhead{}
\renewcommand{\headrulewidth}{0pt} 
\renewcommand{\footrulewidth}{0pt}
\setlength{\arrayrulewidth}{1pt}
\setlength{\columnsep}{6.5mm}

\makeatletter 
\newlength{\figrulesep} 
\setlength{\figrulesep}{0.5\textfloatsep} 

\newcommand{\topfigrule}{\vspace*{-1pt}%
\noindent{\color{cream}\rule[-\figrulesep]{\columnwidth}{1.5pt}} }

\newcommand{\botfigrule}{\vspace*{-2pt}%
\noindent{\color{cream}\rule[\figrulesep]{\columnwidth}{1.5pt}} }

\newcommand{\dblfigrule}{\vspace*{-1pt}%
\noindent{\color{cream}\rule[-\figrulesep]{\textwidth}{1.5pt}} }

\makeatother

\twocolumn[
  \begin{@twocolumnfalse}{}

\par
\vspace{1em}
\sffamily
\begin{tabular}{m{0.5cm} p{16.5cm} }

 & \noindent\LARGE{\textbf{Time Response of Water-based Liquid Scintillator from X-ray Excitation}} \\
\vspace{0.3cm} & \vspace{0.3cm} \\

 & \noindent\large{Drew R. Onken,\textit{$^{a}$} Federico Moretti,\textit{$^{a}$} Javier Caravaca,\textit{$^{a,b}$} Minfang Yeh,\textit{$^{c}$} Gabriel D. Orebi Gann,\textit{$^{a,b}$} and Edith D. Bourret\textit{$^{a}$}} \\
\vspace{0.3cm} & \vspace{0.3cm} \\
  & \noindent\normalsize{Water-based liquid scintillators (WbLS) present an attractive target medium for large-scale detectors with the ability to enhance the separation of Cherenkov and scintillation signals from a single target. This work characterizes the scintillation properties of WbLS samples based on LAB/PPO liquid scintillator (LS). X-ray luminescence spectra, decay profiles, and relative light yields are measured for WbLS of varying LS concentration as well as for pure LS with a range of PPO concentrations up to 90 g/L. The scintillation properties of the WbLS are related to the precursor LAB/PPO: starting from 90 g/L PPO in LAB before synthesis, the resulting WbLS have spectroscopic properties that instead match 10 g/L PPO in LAB. This could indicate that the concentration of active PPO in the WbLS samples depends on their processing.} \\
\vspace{0.3cm} & \vspace{0.3cm} \\

\end{tabular}

 \end{@twocolumnfalse} \vspace{0.6cm}

  ]

\renewcommand*\rmdefault{bch}\normalfont\upshape
\rmfamily
\section*{}
\vspace{-1cm}


\footnotetext{\textit{$^{a}$~Lawrence Berkeley National Laboratory, Berkeley, CA 94720, USA.}}
\footnotetext{\textit{$^{b}$~University of California, Berkeley, CA 94720, USA. }}
\footnotetext{\textit{$^{c}$~Brookhaven National Laboratory, Upton, NY 11973, USA. }}




\section{Introduction}
\label{sec:intro}

The ability to detect Cherenkov and scintillation signals from a single target has many applications across particle and nuclear physics, nuclear nonproliferation, and medical physics. The resulting particle identification capabilities, combined with directional reconstruction at low energies, offer the potential for unprecedented levels of event discrimination and background rejection.  Use of a novel water-based liquid scintillator (WbLS) medium is one route by which this capability could be realized \cite{Yeh2011,Alonso2014}, and has been proposed for use in the THEIA~\cite{Alonso2014,Askins2019}, ANNIE~\cite{Back2019}, and WATCHMAN~\cite{Askins2015} experiments.

One liquid scintillator (LS) commonly used in particle physics experiments is linear alkylbenzene (LAB) \cite{Beriguete2014,Andringa2016,Park2013,Yu2016}. Often used with the fluor 2,5-diphenyloxazole (PPO), this LAB/PPO system has been extensively characterized~\cite{Xiao2010,Kogler2013,Okeeffe2011}. Studies have been conducted on the impact of varying the PPO concentration on the emission spectra and the luminescence decay profile~\cite{Li2011,Lombardi2013,Undagoitia2009,Ye2015}. Furthermore, the separation of the Cherenkov signal from a scintillating target has been demonstrated in LAB~\cite{Li2016,Gruszko2019} and also in LAB/PPO~\cite{Caravaca2017a,Kaptanoglu2019}. This separation in the LAB/PPO is challenging because the addition of PPO both increases the scintillation light yield by an order of magnitude, swamping the Cherenkov component, and shortens the fast scintillation time constant, making it hard to separate the two light sources in time.

WbLS offers enhanced light production relative to pure water, reduced reabsorption relative to pure liquid scintillator (LS), and the ability to tune the relative ratio of the Cherenkov and scintillation signals based on the fractional scintillator content. The emission spectrum, timing, and light yield can be modified by the addition of fluors. A promising WbLS cocktail based on LAB/PPO has been synthesized at Brookhaven National Laboratory (BNL) \cite{Yeh2011}. A few studies have begun to examine the scintillator properties of the WbLS, examining how these mixtures perform and how they differ from the well-characterized pure LAB/PPO scintillators~\cite{So2014,Bignell2015b,Bignell2015a}. However, further characterization is still needed, especially on the WbLS luminescence time response. 

This work investigates the scintillation properties of WbLS with varying concentration of LAB/PPO (LS) dispersed in water, focusing on the X-ray excited luminescence spectra, time decay, and light yield. For comparison, similar measurements are conducted on a series of pure (water-free) LAB/PPO samples with varying PPO concentrations up to 90 g/L (the concentration used as the LS precursor for WbLS synthesis here). None of the prior works have examined PPO concentrations in LAB above 10 g/L. Together, this study examines how WbLS differs from the LAB/PPO precursor, informing how these WbLS samples may be further tailored to optimize light yield or decay time for specific applications.

\section{Methods}

\subsection{WbLS Samples}

WbLS is a mixture of water and an organic oil-based scintillator, combined using surfactants. The WbLS samples used in this study were synthesized at BNL. These samples contained 1\%, 5\%, or 10\% LS in water, each using a LS precursor made of 90 g/L PPO in LAB. Such a high concentration of PPO was used in the precursor to ensure enough fluor in the resulting WbLS. For comparison, we also prepared a series of pure (water-free and surfactant-free) LS samples with a range of PPO concentrations increasing up to 90 g/L in LAB. 

\subsection{Scintillation Characterization} 

All liquid scintillator samples were measured in sealed quartz tubes with 4 mm inner diameter. Steady-state X-ray luminescence measurements were obtained using a Bruker 50 kV (60 mA) rotating copper anode X-ray generator, directed into the sample perpendicular to the collection optics. These X-ray energies are well below the threshold for Cherenkov radiation; this work focuses exclusively on scintillation light. Emission spectra were obtained using a SpectraPro-2150i spectrometer coupled to a PIXIS:100B charge-coupled detector, with a spectral correction applied. 

Time-dependent X-ray luminescence was measured using time-correlated single-photon counting. Detailed further in references~\cite{Derenzo2000,Derenzo2008}, this 40 kV pulsed X-ray source is driven by 200 fs Nd:YAG laser pulses and has an impulse response of 100 ps FWHM. Luminescence is detected by a Hamamatsu R3809U microchannel PMT and is processed through an Ortec 9308 picosecond analyzer. For these samples, a laser frequency of --Hz was used to measure decay profiles out from 100 ns before excitation to 550 ns after excitation. 

To help quantify the differences in decay profiles, a multi-exponential function is fit to each profile. Component fitting of the rise and decay uses a sum of exponential decay functions convoluted with the impulse response \cite{Derenzo2008}. For the data presented in this work, a 4-component fit (one rise component and three decay components, plus a constant fraction) was deemed to be optimal for capturing the features of the decay and approaching the best fit (minimizing $\chi^2$) without using too many components. It is important to note that there is not a physical basis for all four exponential components in this fit; this fit is a tool to quantify differences in the decay profiles.

\begin{figure}
	\centering
	\includegraphics[width=\columnwidth]{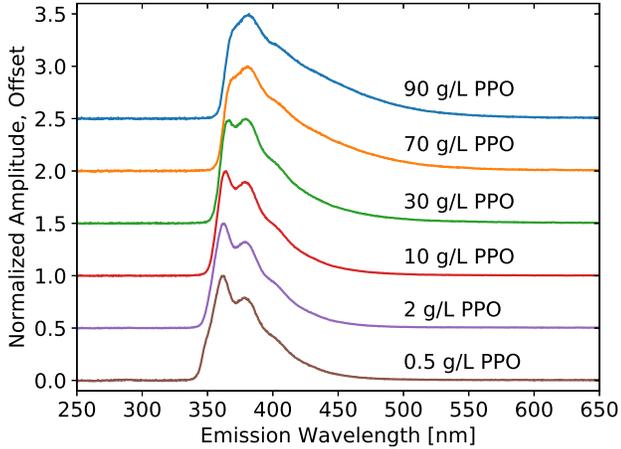}
	\caption{Emission spectra resulting from X-ray excitation of pure LS with varying concentrations of PPO in LAB. Each curve is displayed normalized at its maximum and offset along the vertical axis.}
	\label{f:ppoSeries_Xemission}
\end{figure}

\begin{figure}
	\centering
	\includegraphics[width=\columnwidth]{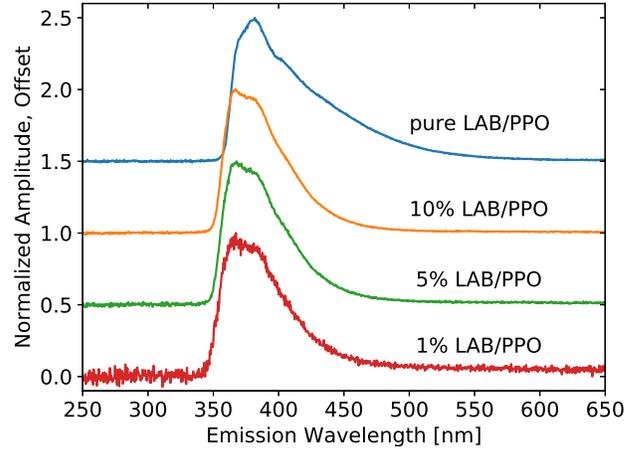}
	\caption{Emission spectra resulting from X-ray excitation of pure LS (LAB with 90 g/L PPO) and the three WbLS concentrations made from this LS. Each curve is displayed normalized at its maximum and offset along the vertical axis.}
	\label{f:labppo_Xemission}
\end{figure}

Relative light yield is measured by integrating the light in the decay profiles from the time-correlated single photon counting measurements, accounting for any variance in exposure time and intensity. These values are reported relative to a standard EJ-301 liquid scintillator \cite{EljenTechnology} also measured by this system. To account for the difference in X-ray absorption between the LAB/PPO and water, a correction factor was generated using a Geant4-based Monte Carlo model that accounts for the composition of the different materials. According to our model, the deposited energy from our 40 kV X-ray source in LAB/PPO is about 65\% of that deposited in WbLS. Although the luminescence spectra vary in shape, these spectral changes are minor considering the rather flat response of the PMT. 


\section{Results and Discussion}\label{s:res}

X-ray luminescence spectra were measured on the pure LS compounds with a range of PPO concentrations in LAB, as shown in Figure \ref{f:ppoSeries_Xemission}. The PPO emission bands are noticeable at each concentration, though their relative intensity is modified noticeably with concentration, due to reabsorption. As the concentration is increased, the 345 nm component shoulder becomes no longer visible and the low-energy emission edge shifts towards higher wavelengths. Increasing concentration further, the 360 nm component becomes weaker relative to the 380 nm component. The same measurement is shown in Figure \ref{f:labppo_Xemission} for the three samples of WbLS (1\%, 5\%, and 10\% LS content). The pure LS precursor (90 g/L PPO in LAB) is reproduced in this same plot for reference. While all three water-based compounds appear to have identical emission spectra, they clearly differ from that of their pure LS precursor. Instead, the relative intensity of the 380 nm component versus the 360 nm component is much more similar to that of the 10 g/L PPO sample, though the WbLS components are less resolved. 

\begin{figure}
	\centering
	\includegraphics[width=\columnwidth]{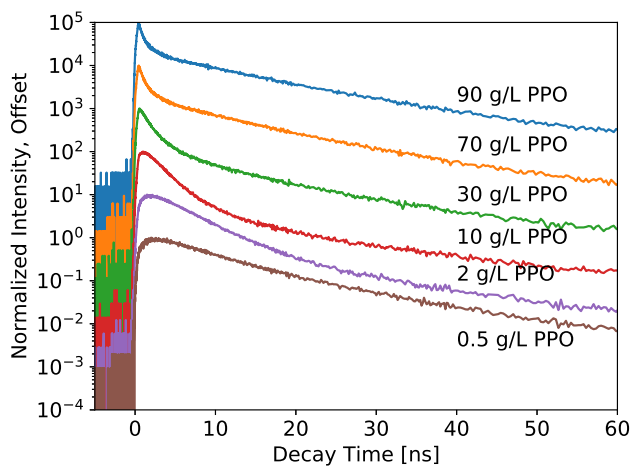}
	\caption{Time profiles of pure LS for varying concentrations of PPO in LAB from pulsed X-ray excitation. Each curve is normalized by its maximum and then is scaled by a power of 10 to offset along the vertical axis, in order to more clearly show profile shape differences. }
	\label{f:labppo_time}
\end{figure}

\begin{figure}[t!]
	\centering
	\includegraphics[width=\columnwidth]{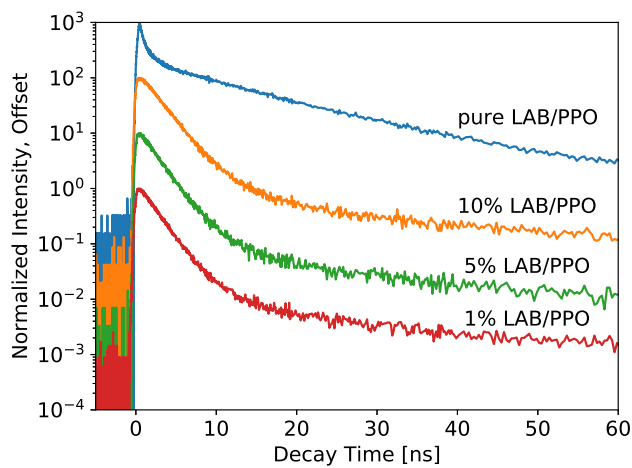}
	\caption{Time profiles of pure LS (90 g/L PPO in LAB) and the three WbLS concentrations from pulsed X-ray excitation. Each curve is normalized by its maximum and then is scaled by a power of 10 to offset along the vertical axis, in order to more clearly show profile shape differences. }
	\label{f:wbls_time}
\end{figure}

\begin{figure}[t!]
	\centering
	\includegraphics[width=1\columnwidth]{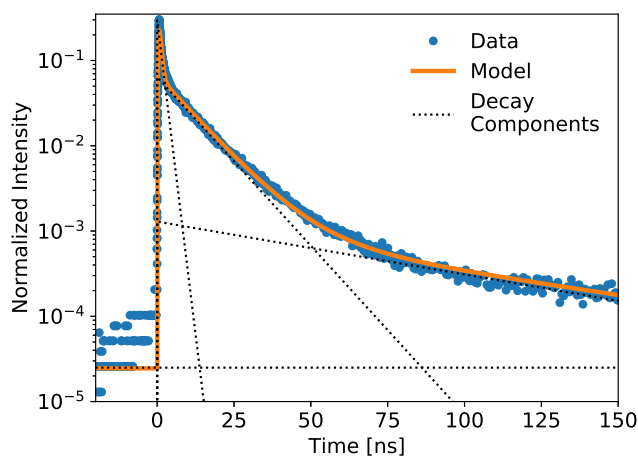}
	\caption{Showing the components of a 4-exponential fitting (1 rise, 3 decay, plus constant) of the pure LS precursor (90 g/L PPO in LAB) response to a pulsed X-ray source. The data is normalized by integration. Decay lifetimes and weight fractions for the component fitting of all samples are displayed in Tables \ref{t:decayComponentPPO} and \ref{t:decayComponent}. Data collected out to +550 ns were used to fit these decay components. \color{red} \textbf{ }}
	\label{f:wbls_component}
\end{figure}

The luminescence decay profiles measured under X-ray excitation are shown in Figures \ref{f:labppo_time} and \ref{f:wbls_time}. Decay times are determined from 4-exponential fitting, as demonstrated in Figure \ref{f:wbls_component}. As the concentration of PPO increases from 0.5 g/L to 90 g/L in pure LS (Figure \ref{f:labppo_time}), the decay time of the fastest decay component shortens dramatically from 6 ns down to 0.5 ns. For most samples, the fast decay component accounts for the highest luminescence fraction. Only at the two highest PPO concentrations does the second decay component overtake the fastest component as a more significant contributor to the total luminescence (Table \ref{t:decayComponentPPO}). These values and trends are very similar to those reported in the literature: the primary decay components measured in this work under X-ray excitation are slightly slower than those measured under fluorescence \cite{Li2011,Lombardi2013}, match the oxygenated electron response in ref. \cite{Okeeffe2011}, and are slightly faster than those measured under $\alpha$ and $\beta$ irradiation \cite{Lombardi2013} and $\gamma$ irradiation \cite{Undagoitia2009}, for the corresponding concentrations of PPO. 

For the WbLS decay profiles (Figure \ref{f:wbls_time}), decay time fits are shown in Table \ref{t:decayComponent}. All three WbLS samples behave similarly, with their primary decay component around 2 ns (for 87\% of total luminescence decay). This stands in contrast to the pure 90 g/L PPO sample, which has a weaker but faster initial decay of 0.5 ns (36\%). The decay profile for the WbLS samples more closely matches that of the 10 g/L PPO, with a primary decay time of 2.0 ns (79\%). PPO is insoluble in water, but even in the presence of the surfactant, it is plausible that some PPO is not incorporated in the micelles with the LAB. This would reduce the concentration of PPO in the micelles and explain why, from a spectroscopic point of view, the WbLS synthesized from 90 g/L PPO appears to be identical to 10 g/L PPO. It is suspected that this is due to the WbLS processing, specifically on the role of the surfactant in the micelle formation, and needs to be further studied.

While it seems that for the WbLS, the concentration of PPO in the LAB micelles is reduced from that in the precursor solution, this effect is not linear with the amount of water added. With water added to the LS using surfactants to make the 1\%, 5\%, and 10\% WbLS samples, the average concentration of PPO \textit{in the total volume} becomes 0.9 g/L, 4.5 g/L, and 9.0 g/L, respectively. Yet the decay profile for all three WbLS samples lines up most closely with the pure 10 g/L PPO in LAB sample. Average concentration of PPO in the water-based samples does not impact luminescence decay time; rather it is the concentration of PPO remaining in the scintillating LAB/PPO micelles that matters, and this amount appears to only scale weakly with emulsion in water below 10\% LS content.

\begin{table*}
    \centering
    \begin{tabular}{c||c||c|c||c|c||c|c}
      LS Samples & $\tau_{rise}$ [ns] & $\tau_1$ [ns] & $f_1$ [\%] & $\tau_2$ [ns] & $f_2$ [\%] & $\tau_3$ [ns] & $f_3$ [\%] \\\hline
0.5 g/L PPO & 0.98 $\pm$ 0.04 & 7.0 $\pm$ 0.6 & 61 & 16. $\pm$ 1. & 34 & 80 $\pm$ 5 & 4.4\\
1 g/L PPO & 0.92 $\pm$ 0.07 & 5.5 $\pm$ 0.3 & 72 & 13.7 $\pm$ 0.9 & 25 & 84 $\pm$ 2 & 3.7\\
2 g/L PPO & 0.88 $\pm$ 0.05 & 4.3 $\pm$ 0.2 & 82 & 13.4 $\pm$ 0.8 & 15 & 86 $\pm$ 1 & 3.7\\
3 g/L PPO & 0.83 $\pm$ 0.06 & 3.6 $\pm$ 0.1 & 84 & 13.4 $\pm$ 0.9 & 12 & 83 $\pm$ 2 & 4.0\\
5 g/L PPO & 0.69 $\pm$ 0.03 & 2.81 $\pm$ 0.07 & 84 & 13.0 $\pm$ 0.8 & 12 & 80 $\pm$ 2 & 4.4\\
10 g/L PPO & 0.49 $\pm$ 0.03 & 2.06 $\pm$ 0.04 & 77 & 11.2 $\pm$ 0.4 & 18 & 72 $\pm$ 2 & 5.1\\
30 g/L PPO & 0.33 $\pm$ 0.12 & 1.15 $\pm$ 0.02 & 55 &  9.9 $\pm$ 0.5 & 38 & 64 $\pm$ 4 & 7.0\\
70 g/L PPO & 0.26 $\pm$ 0.03 & 0.58 $\pm$ 0.02 & 38 & 10.5 $\pm$ 0.4 & 55 & 69 $\pm$ 4 & 7.1\\
90 g/L PPO & 0.24 $\pm$ 0.02 & 0.48 $\pm$ 0.01 & 35 & 11.1 $\pm$ 0.1 & 58 & 72 $\pm$ 3 & 7.4\\
    \end{tabular}
    \caption{Rise times ($\tau_{rise}$ in ns), decay times ($\tau_i$ in ns), and decay component fractions ($f_i$ in \%) are displayed for each pure LS sample, varying PPO concentration in LAB, from excitation with a pulsed X-ray excitation source, as shown in Figures \ref{f:labppo_time} and \ref{f:wbls_component}. The error bounds listed are 95\% confidence intervals measured by repeating this experiment; thus they account for any random error of sample positioning, data acquisition, and component fitting.    }
    \label{t:decayComponentPPO}
\end{table*}

\begin{table*}
    \centering
    \begin{tabular}{c||c||c|c||c|c||c|c}
      WbLS Samples & $\tau_{rise}$ [ns] & $\tau_1$ [ns] & $f_1$ [\%] & $\tau_2$ [ns] & $f_2$ [\%] & $\tau_3$ [ns] & $f_3$ [\%] \\\hline
1\% LAB/PPO & 0.23 $\pm$ 0.06  & 2.00 $\pm$ 0.03 & 87 & 12 $\pm$ 1 & 6.8 & 110 $\pm$ 10 & 6.2\\
5\% LAB/PPO & 0.23 $\pm$ 0.04 & 2.00 $\pm$ 0.02 & 88 & 10.0 $\pm$ 0.6 & 6.6 & 106 $\pm$ 6  & 5.7\\
10\% LAB/PPO & 0.29 $\pm$ 0.03 & 2.22 $\pm$ 0.03 & 89 & 10.7 $\pm$ 0.9 & 6.0 & 102 $\pm$ 9  & 5.5\\
    \end{tabular}
    \caption{Rise times ($\tau_{rise}$ in ns), decay times ($\tau_i$ in ns), and decay component fractions ($f_i$ in \%) are displayed for each WbLS sample, from excitation with a pulsed X-ray excitation source, as shown in Figure \ref{f:wbls_time}. }
    \label{t:decayComponent}
\end{table*}

\begin{figure}
	\centering
	\includegraphics[width=\columnwidth]{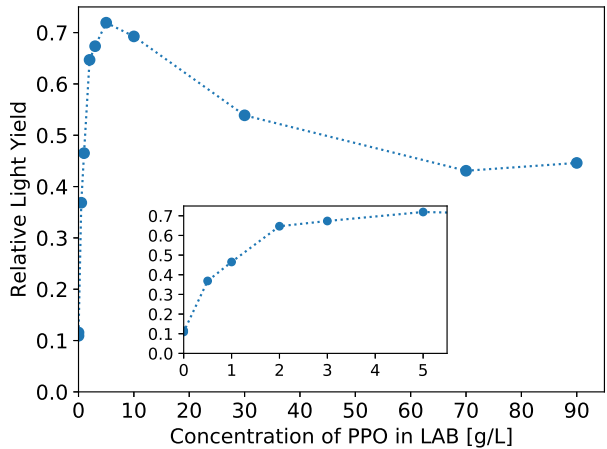}
	\caption{Relative light yield for the pure LS samples, varying as a function of PPO concentration in LAB. Light yield values are presented as a fraction of the EJ-301 light yield. The inset focuses on concentrations $\leq$5 g/L PPO.}
	\label{f:ly_labppo}
\end{figure}

\begin{figure}
	\centering
	\includegraphics[width=\columnwidth]{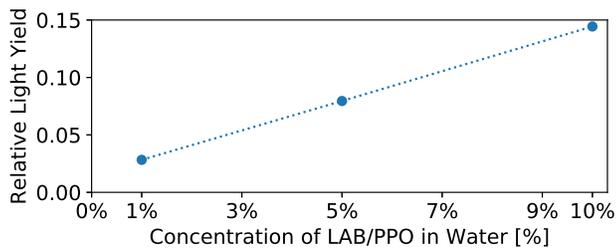}
	\caption{Relative light yield for the WbLS samples, varying as a function of concentration of LS in water. Each sample was synthesized from a 90 g/L PPO in LAB precursor. Light yield values are presented as a fraction of the EJ-301 light yield. Correction factors have been applied to account for the different X-ray absorption cross-sections between LAB/PPO and water. }
	\label{f:ly_wbls}
\end{figure}

The light yield of the scintillators is reported as a fraction relative to an EJ-301 standard. Figure \ref{f:ly_labppo} shows the relative light yield of the pure LS as a function of PPO concentration. For these samples measured in our small 4 mm cuvettes, the 5 g/L PPO sample gives the highest light yield. This agrees with other works which report relative light yield reaching a maximum around this same concentration \cite{Lombardi2013}, even in larger 250 mm vessels \cite{Ye2015}, though neither reference tests PPO concentrations above 7 g/L. Figure \ref{f:ly_wbls} shows the light yield for the WbLS samples varying with the percentage of LS in water. The light yield increases with the concentration of micelles containing LAB/PPO.

\section{Conclusions}\label{s:conc}

This work provides decay profile measurements of WbLS, with comparisons to pure LAB/PPO with different PPO concentrations up to 90 g/L. WbLS emulsions of 1\%, 5\%, and 10\% LAB/PPO all have similar luminescence spectra and decay components. Comparing the spectroscopic properties, the WbLS samples appear more similar to the water-free 10 g/L PPO in LAB sample than to the water-free 90 g/L PPO in LAB sample that served as the precursor. This suggests that during the synthesis process, a reduced amount of PPO is incorporated into the LAB micelles. With this knowledge, the change in PPO concentration could be taken into account during the synthesis process. Future WbLS compounds can also be optimized for specific applications using the measurements in this work: PPO concentration can be selected to tune the scintillation decay time and light yield. Specifically, for a WbLS target medium designed to enhance the separation of Cherenkov and scintillation light, a lower PPO concentration (e.g. targeting 1 g/L) could be used to further slow the scintillation decay time (from 2.0 to 5.5 ns) and reduce the light yield (by a factor of 1.5), thereby increasing separation from the fast Cherenkov radiation.

\section*{Conflicts of interest}
There are no conflicts to declare.

\section*{Acknowledgements}
The authors would like to thank Benjamin J. Land for his contributions in early discussions of these measurements. This work was performed under the auspices of the U.S. Department of Energy by Lawrence Berkeley National Laboratory under Contract DE-AC02-05CH11231. The project was funded by the U.S. Department of Energy, National Nuclear Security Administration, Office of Defense Nuclear Nonproliferation Research and Development (DNN R\&D). The authors from UC-Berkeley were also supported by the U.S. Department of Energy, Office of Science, Office of High Energy Physics, under Award Number DE-SC0018974. The work conducted at Brookhaven National Laboratory was supported by the U.S. Department of Energy under Contract DE-AC02-98CH10886.






\end{document}